%% file: main.tex
\documentclass[journal]{IEEEtran}
\ifCLASSINFOpdf
\else
\fi

\usepackage{xcolor}
\usepackage{amsmath}
\usepackage{bm}
\usepackage{graphicx}
\usepackage{multirow}
\usepackage{threeparttable}
\usepackage{lipsum}
\usepackage{mathtools}
\usepackage{cuted}
\usepackage{cite}
\newcommand{\cmmnt}[1]{}
\usepackage[nomargin,inline,index]{fixme}
\fxsetup{status=draft,theme=color}
\definecolor{fxwarning}{rgb}{1.0000,0.0000,0.0000}

\DeclareUnicodeCharacter{2212}{-}

\begin{document}

\title{Multi-Wavelength Photonic Neuromorphic Computing for Intra and Inter-Channel Distortion Compensations in WDM Optical Communication Systems}

\author{Benshan~Wang,
        ~Thomas~Ferreira~de~Lima, 
       ~Bhavin J Shastri,~\IEEEmembership{Senior Member,~IEEE},
        ~Paul R Prucnal,~\IEEEmembership{Life~Fellow,~IEEE},
        ~Chaoran~Huang,~~\IEEEmembership{Member,~IEEE}

\thanks{B. Wang, C. Huang are with the Department of Electronic Engineering, the Chinese University of Hong Kong, Shatin, Hong Kong. e-mail: crhuang@ee.cuhk.edu.hk.}

\thanks{B. Shastri with the Department of Physics, Engineering Physics $\&$ Astronomy, Queen’s University, Kingston, ON KL7 3N6, Canada.}
\thanks{T. Ferreira de Lima and P. Prucnal are with the Department of Electrical and Computer Engineering, Princeton University, Princeton, NJ, 08540.}
\thanks{T. Ferreira de Lima is now with NEC Laboratories America, Princeton, 08540, New Jersey, USA}
\thanks{Manuscript received April 19, 2005; revised August 26, 2015.}}

\markboth{Journal of \LaTeX\ Class Files,~Vol.~14, No.~8, August~2015}%
{Shell \MakeLowercase{\textit{et al.}}: Bare Demo of IEEEtran.cls for IEEE Journals}

\maketitle

\begin{abstract}

DSP (digital signal processing) has been widely applied in optical communication systems to mitigate various signal distortions and has become one of the key technologies that have sustained data traffic growth over the past decade. However, the strict energy budget of application-specific integrated circuit-based DSP chips has prevented the deployment of some powerful but computationally costly DSP algorithms in real applications. As a result, fiber nonlinearity-induced signal distortions impede fiber communications systems, especially in wavelength-division multiplexed (WDM) transmission systems. To solve these challenges in DSP, there has been a surge of interest in implementing neural networks-based signal processing using photonics hardware (i.e., photonic neural networks). Photonic neural networks promise to break performance limitations in electronics and gain advantages in bandwidth, latency, and power consumption in solving intellectual tasks that are unreachable by conventional digital electronic platforms. This work proposes a photonic recurrent neural network (RNN) capable of simultaneously resolving dispersion and both intra- and inter-channel fiber nonlinearities in multiple WDM channels in the photonic domain, for the first time to our best knowledge. Furthermore, our photonic RNN can directly process optical WDM signals in the photonic domain, avoiding prohibitive energy consumption and speed overhead in analog to digital converters (ADC). Our proposed photonic RNN is fully compatible with mature silicon photonic fabrications. We demonstrate in simulation that our photonic RNN can process multiple WDM channels simultaneously and achieve a reduced bit error rate compared to typical DSP algorithms for all WDM channels in a pulse-amplitude modulation 4-level (PAM4) transmission system, thanks to its unique capability to address inter-channel fiber nonlinearities. In addition to signal quality performance, the proposed system also promises to significantly reduce the power consumption and the latency compared to the state-of-the-art DSP chips, according to our power and latency analysis. 

\end{abstract}


\IEEEpeerreviewmaketitle

\section{Introduction}

\IEEEPARstart{G}{lobal} internet traffic is growing exponentially, as driven by bandwidth-hungry applications, such as cloud computing, virtual/augmented reality, and high-definition video streaming~\cite{cisco}. To meet the demands for capacity growth, today’s optical communication systems widely deploy densely spaced wavelength-division multiplexed (WDM) channels and high-order modulation formats~\cite{winzer2008advanced}, which inevitably leads to increased signal distortions. The signal distortions originate from the combined effects of chromatic dispersion (CD) and Kerr nonlinearity in optical fibers. In WDM systems, Kerr nonlinearity causes more severe impairments due to the cross-phase modulation (XPM) and four-wave mixing (FWM) among different wavelength channels. 

Digital signal processing (DSP) technology is widely employed in today’s transport systems to compensate for signal distortions. It has successfully supported fiber capacity growth over the past decade~\cite{kikuchi2015fundamentals,ip08JLT,ip2008coherent,savory2008digital}. However, in handling the data rates in state-of-the-art optical communication systems, the power dissipation of 7 nm complementary metal-oxide-semiconductor (CMOS) DSP chips for 800-Gigabit Ethernet has already approached the maximum thermal dissipation capacity of modern packaging technologies~\cite{QSFPDDWhitepaper,valavala2018thermal}. The problem of surpassing the thermal capacity has necessitated a careful compromise in the complexity of DSP algorithms to ensure that power consumption is within an acceptable limit. As a result, some powerful but computationally intensive DSP algorithms have not been employed in practice. A prominent algorithm for fiber nonlinearity compensation is the digital backpropagation (DBP) algorithm~\cite{ip08JLT,mateo10OE,dar14JLT,tang21OE,zhang22OFC,inoue22OE}, which is capable of solving both intra- and inter-channel distortions by inversing the coupled multi-channel nonlinear Schrödinger equation using the received WDM signals as the inputs to the equation. However, the consequence is that DBP must process several high-speed and large bandwidth WDM channels concurrently, resulting in prohibitively large bandwidth and data communication overhead while being too complicated to implement on an application-specific integrated circuits (ASIC) chip~\cite{zhang2019field}. As a result, nonlinear distortions in fiber communications systems remain an impediment, particularly in WDM transmission systems.

More fundamentally, a critical aspect of increasing fiber capacity relies on consistently reducing the power consumption per bit of the DSP chip as the data rate increases. A survey article by Frey \textit{et al.}~\cite{frey17VDE} compared the scaling of integrated circuit node size and DSP ASIC for optical communications. The result showed that DSP ASICs closely follow the current CMOS generations. This implies that increases in the fiber capacity over the past decade heavily rely on the node size scaling promised by Moore’s Law. However, CMOS node scaling will inevitably slow down and finally halt~\cite{shalf20PTRSA,waldrop16Nature}, suggesting that signal processing hardware needs to be redefined to maintain the explosive internet traffic growth in the future. 

The field of neuromorphic photonics aims to build practical photonic neural networks (PNNs) to solve these challenges in DSP~\cite{tait14JLT, Shen17NP,cheng17SA, tait17SR, lin18Science,feldmann19Nature,Feldmann21Nature,shastri21NP,huang22APX}. PNNs emulate the biological or artificial neural network models using high-speed devices and large-bandwidth photonic waveguides, thus allowing PNNs to execute neural network algorithms (e.g., deep learning) with unmatched speed. PNNs, like other machine learning algorithms, can learn optical fiber transmission channel characteristics from abundant transmission data without explicit channel knowledge and then invert various transmission impairments using the learned transmission channel model~\cite{Huang:21,huang22APX}. However, unlike digital platforms such as ASICs, which heavily rely on parallel computing to keep up with the ever-increasing data rate (resulting in significant power consumption overhead), PNNs leverage photonic devices that are originally designed for optical communications, and thus, if appropriately designed, can always offer a processing speed matching the fiber communication line rate in the future.

Photonic reservoir computing (RC)~\cite{appeltant11NC,vandoorne14NC,vinckier15Optica,larger17PRX,nakajima21CP}, a subclass of recurrent neural networks (RNNs), was first proposed and demonstrated to process optical communication systems fast and efficiently~\cite{antonik16TNNLS,argyris18SR, Coarer18JSTQE, argyris19JSTQE,daros20JSTQE, sackesyn21OE, donati22OE}. RC consists a reservoir of randomly linked neurons followed by a readout layer. The weight connections are fixed inside the reservoir. Only the readout layer's weights are trained using linear regression. Photonic RCs for optical communications have attracted significant research interest due to their ease of implementation. They have been implemented using both optical fibers and integrated photonic platforms, and have demonstrated many functions, including dispersion and nonlinear compensations for both IM/DD and coherent optical systems. The reader is directed to the most recent review articles in this field~\cite{nakajima21reservoir,brunner2018tutorial}. Since only the readout layer of RC can be trained, RC faces the challenges of limited expressivity and thus sometimes cannot guarantee convergence to desired behaviors. 

As proposed by some authors~\cite{Lima20Nano}, a different method of overcoming RC's constraint is creating a rigorous mapping between physical models of optoelectronic systems and abstract models of neural networks. Consequently, PNNs can then use standard machine learning methods (such as back-propagation) to train every parameter in the PNN. We used this approach to demonstrate a silicon photonic feed-forward neural network (FNN) to model and compensate for fiber nonlinearities in a 10,080 km trans-pacific transmission system~\cite{huang21NE}. The PNN functions as a photonic accelerator to replace the power-intensive nonlinear compensation module in DSP. However, the power analysis reveals that the digital to analog converters (DAC) used to convert digital signals to analog photonic neural network inputs contribute to most of the power consumption in the system, undermining the benefits of PNN in terms of energy efficiency~\cite{huang22APX}.

\begin{figure*}[ht]
\centering
\includegraphics[width=\linewidth]{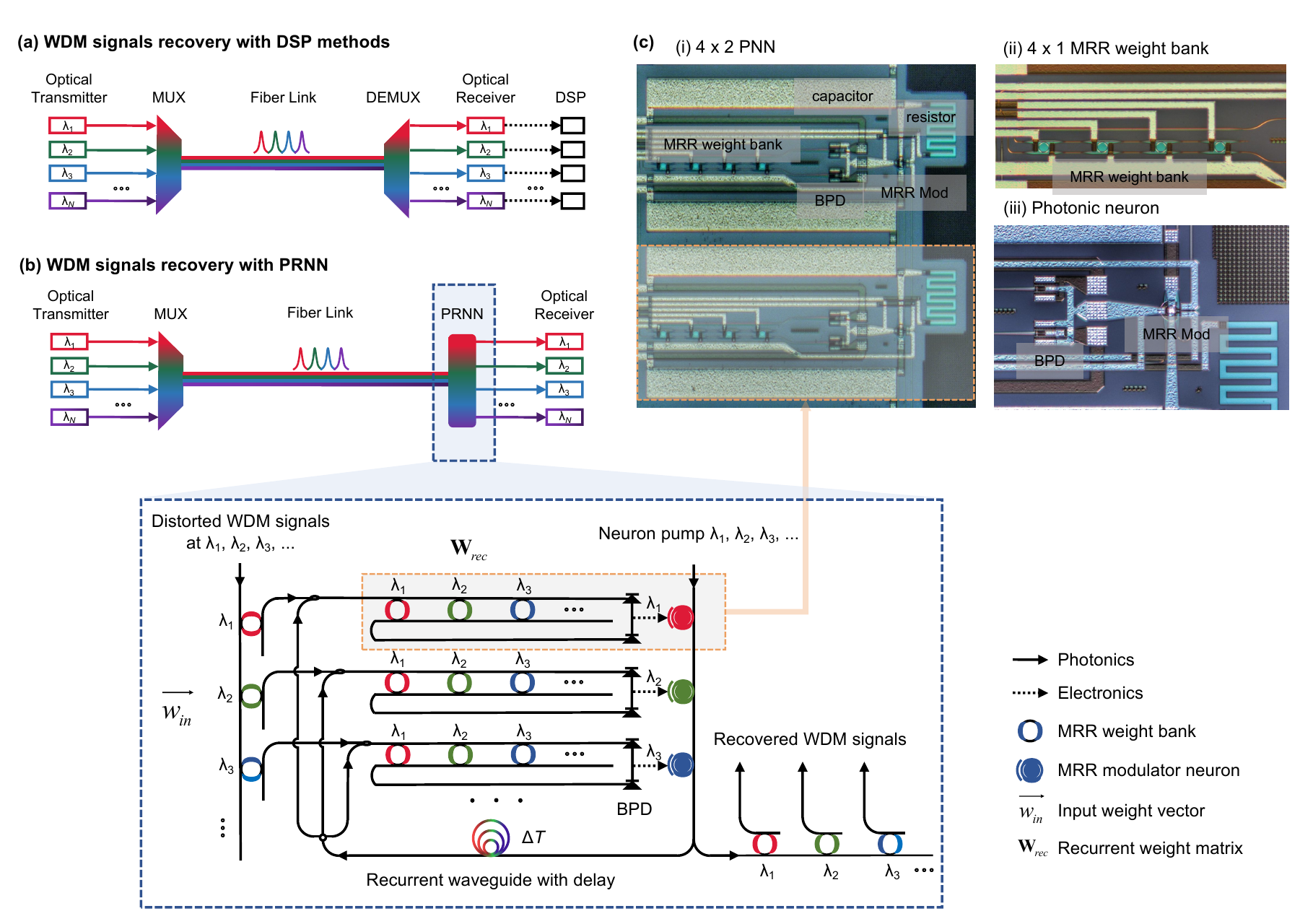}
\caption{ (a) Conventional WDM transmission system with DSP. (b) Proposed WDM transmission system where signal recovery is performed in the optical domain using the proposed photonic recurrent neural network before the optical receiver. (c) A false-colour confocal micrograph of the silicon photonic neuron device. (c) is reproduced from Huang $\textit{et al}$., \textit{Nat. Electron.}~\textbf{28},~8~(2021)~ref.\cite{huang21NE}. WDM: wavelength-division multiplexing, DSP: digital signal processor, (DE)MUX: (de)multiplexing, PRNN: photonic recurrent neural network, MRR: microring resonator, BPD: balanced photodetector.}
\label{fig:system and model}
\end{figure*}

To address these challenges, this paper proposes a WDM-based photonic recurrent neural network (depicted in Figure~\ref{fig:system and model} (b))~\cite{tait17SR} as a front-end processor, capable of simultaneously resolving dispersions and both intra- and inter-channel fiber nonlinearities in multiple WDM channels in the photonic domain, for the first time to our best knowledge.
Furthermore, our photonic RNN can directly interface and process optical WDM signals in the photonic domain, eliminating prohibitive energy consumption overhead and speed reduction in ADCs. The WDM-based photonic neural network encodes the information to different wavelengths as inputs, and thus it can directly interface with the WDM optical communication systems. The proposed photonic neural network composes microring resonator (MRR) banks for synaptic weighting, photodetector-modulator neurons as activation function, and a feedback waveguide providing all-to-all recurrency connecting all WDM channels. The proposed photonic RNN can be analogous to optical fiber transmission systems in that their linear neuron-to-neuron connections with the internal feedback waveguide are analogous to dispersive memory. Meanwhile, the neuron nonlinearity can mimic all forms of nonlinear effects (not limited to fiber nonlinearities) in a fiber transmission system~\cite{Huang:21}. Thanks to these analogies, the photonic RNN can be trained to resemble the fiber transmission channels and invert the various channel impairments. More interestingly, PNN provides wideband weighted addition among WDM channels, allowing addressing inter-channel nonlinearity problems in WDM systems that DSP built on ASIC chips lacks the bandwidth or power to process.

To evaluate performance of our proposed front-end neuromorphic processor, we apply the proposed photonic RNN to a WDM pulse-amplitude modulation 4-level (PAM4) system and evaluate the signals’ bit error rate (BER) reduction after being processed by the photonic RNN in the numerically simulated transmission system. We compare the signal performances with various DSP approaches, demonstrating that our photonic RNN has smaller BERs for all wavelength channels. We verify that the performance improvement comes from the unique capability of compensating for inter-channel nonlinearities. Our power consumption and latency analysis show that our photonic RNN can significantly reduce the power consumption and latency compared to the state-of-the-art DSP chips. We hope our proposed photonic RNN, which is fully compatible with commercial silicon photonic platforms, can provide and inspire new solutions to the future high-speed low-energy fiber communication systems.

\section{Principles and System Description}
\label{sec: Principles and System Description}

Figure~\ref{fig:system and model}(a) shows a typical $N$-channel WDM fiber transmission system. At the transmitters, independent information streams are modulated on different wavelengths of lasers and then multiplexed and transmitted simultaneously in a single-mode fiber. The signals experience chromatic dispersion and Kerr effect induced nonlinear distortions in the fiber link as described by the coupled nonlinear Schrödinger equation~\cite{agrawal00NFO} in Equation~\ref{eq: NLSE with FWM and Raman}.

\begin{figure*}[ht]
\centering
\begin{equation}\label{eq: NLSE with FWM and Raman}
\begin{aligned}
    \frac{\partial u_{\lambda_k}}{\partial z}+ \underbrace{i\frac{\beta_2}{2}\frac{\partial^2 u_{\lambda_k}}{\partial t^2}}_{\rm Dispersion}= \underbrace{i\gamma \left| u_{\lambda_k} \right|^2 u_{\lambda_k}}_{\rm Self-phase~modulation} +\underbrace{i2\gamma \sum_{i\neq k}^N \left| u_{\lambda_i} \right|^2 u_{\lambda_k}}_{\rm Cross-phase~modulation}
    + \underbrace{i2\gamma \sum_{k = m + n - l}^N u_{\lambda_m }u_{\lambda_n} u_{\lambda_l}^* {\rm exp}(i \Delta k_{mnlk}z)}_{\rm Four-wave~mixing}\\
    +\underbrace{i\gamma f_R u_{\lambda_k}\sum_{i\neq k}^N\int_{-\infty}^t h_R(t-t')\left[ \left [\left| u_{\lambda_k}(z,t')\right|^2 + \left |u_{\lambda_i}(z,t') \right |^2 \right ] + u_{\lambda_k}(z,t')u_{\lambda_i}^*(z,t'){\rm exp}\left[\pm i\Omega(t-t') \right]\right]{\rm d}t'}_{\rm Stimulated~Raman~scattering}
\end{aligned}
\end{equation}
\end{figure*}
where $u_{\lambda_i}(t,z)$ is the optical field of the signal at wavelength $\lambda_i$, $\beta_2$ is the group-velocity dispersion governing linear impairment, and $\gamma$ is the fiber nonlinear coefficient. $\Delta k_{mnlk}=k_m+k_n-k_l-k_k$ represents the phase mismatch. $f_R$ represents the fractional Raman contribution, $h_R(t)$ is Raman response function for stimulated Raman scattering, $\Omega$ is the Stokes shift~\cite{agrawal00NFO}.
The first term on the right hand side of the NLSE is self-phase modulation which introduces a nonlinear phase related to the signal power, and the second term is cross-phase modulation which describes how the signal in one wavelength is affected by the nonlinear phase introduced by other wavelength signals. The third term is four-wave mixing which involves nonlinear interaction among four optical waves, and the fourth term represents stimulated Raman scattering where one signal acts as a pump and generates the frequency-shift radiation to another signal. The inter-channel stimulated Raman scattering is significant only if in ultra-wide WDM transmission~\cite{semrau18JLT}. The XPM and FWM terms indicate that fully compensating the inter-channel nonlinearity requires the receiver to be able to detect and process multiple wavelength channels simultaneously. At the receiver of a conventional WDM fiber transmission system shown in Figure~\ref{fig:system and model}(a), the WDM signals are first demultiplexed and then detected by the photodetectors and processed by DSPs separately. In principle, the fiber-induced distortions can be compensated by DBP, which solves the $z$-reversed NLSE digitally, leaving only non-deterministic noises\cite{ip08JLT}. However, full compensation of inter-channel effects using DBP is too computationally expensive to implement in DSP because it requires first high oversampling of the signal and then solving a set of coupled NLSE with a fine step size~\cite{mateo2010efficient}. As a result, inter-channel nonlinearity remains one of the major limitations to increasing fiber capacity.

\begin{figure*}[ht]
\centering
\includegraphics[width=\linewidth]{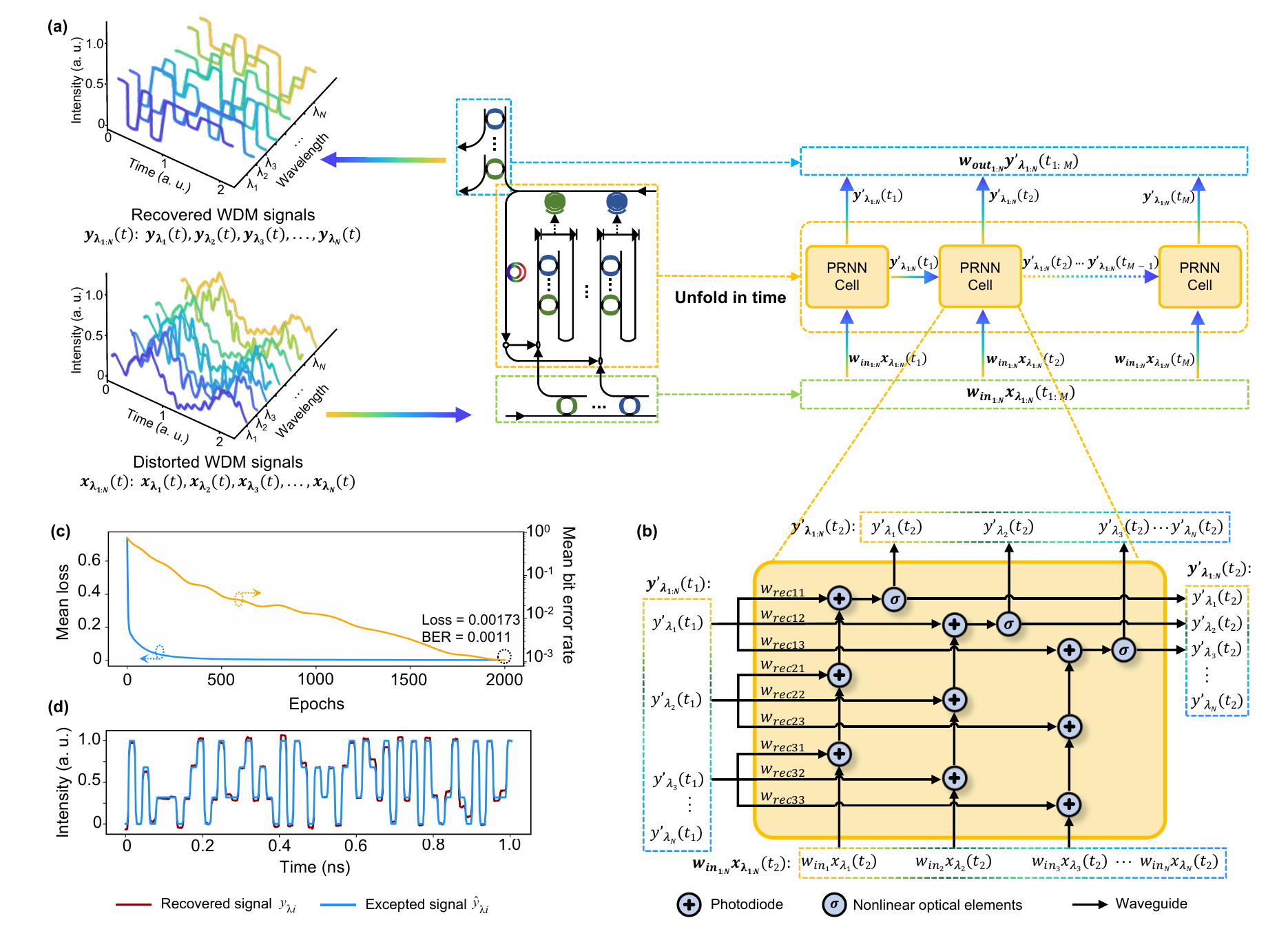}
\caption{Framework and training of the photonic RNN model. (a) Photonic RNN for $N$-channel ($\lambda_1, \lambda_2, \lambda_3,\cdots,\lambda_N$) WDM PAM4 optical signals equalization. The distorted WDM signals $\bm{x_{\lambda_{1:N}}}$ are regarded as inputs of photonic RNN and recovered to original signals $\bm{y_{\lambda_{1:N}}}$. Every channel has the same signal sequence length of $M$. The signals are inputted and outputted via the MRR arrays. $\bm{w_{ini}}, \bm{w_{outi}}$ are the tunable weight vectors for different WDM signals $\lambda_i$ in the input and output MRR array, respectively. Reproduce from Peng \textit{et al.}, \textit{arXiv:2106.13865} (2021) Ref.~\cite{peng21arXiv}. (b) Schematic diagram of the discrete photonic RNN cell. $\sigma$: activation function. (c) The mean training loss and mean BER for 2-channel $\times$ 56 GBaud/s/lane 20 km transmission system. (d) Comparison of the expected signal (blue) and the recovery output (red) of the photonic RNN model.
}
\label{fig:framework}
\end{figure*}

Our proposed photonic RNN solves this problem from a new perspective as shown in Figure~\ref{fig:system and model}(b). The photonic RNN's input is an array of MRRs, each with a reasonance frequency corresponding to the central frequency of the signals in the WDM transmission system. The input MRR array serves as the input layer of the photonic RNN, accepting the WDM signals from the transmission link and providing a tunable weight vector $\bm{w_{in}}$ to the WDM signals at the optical domain. After the input layer, the WDM signals are broadcast to and weighted by the reconfigurable, continuous-valued MRR weight banks~\cite{tait16JSTQE} consisting of $N \times N$ MRRs, where $N$ is the number of WDM channels to be processed. Next, the WDM signals are summed by the followed balanced photodetector which converts the total optical power from each MRR output into photocurrent, resulting in a complementary $-$1 to $+$1 continuous weight range.
The weight matrix given by the $N \times N$ weight bank is denoted as $\bm{W_{rec}}$. The photocurrent modulates the transmission of the MRR modulator (that is, photonic neuron) and hence modulates the optical power of a continuous-wave optical carrier (labeled as a ‘neuron pump’). The MRR modulator exhibits nonlinear electrical-to-optical transfer functions, producing the nonlinear activation function in the photonic neural network. The outputs of the photonic neurons are partially connected back to the $N \times N$ MRRs’ inputs using a single feedback waveguide.

The “Broadcast-and-weight” protocol~\cite{tait14JLT} is compatible with the mainstream silicon photonics platform. Figure~\ref{fig:system and model}(c) shows the micrograph of a $4\times 2$ photonic neural network device fabricated by the commercial silicon photonic foundry as an example, which comprises two arrays of MRR weight banks connected to two photonic neurons (i.e., germanium-silicon photodetector–MRR PN-junction modulator pair). The experimental demonstration of MRR weight banks was first published by Tait \textit{et al.}~\cite{tait14JLT}, followed by several efforts to improve the weighting precision in the network~\cite{tait18OE,ma19OE,huang20APL}. The most recent work demonstrated 9-bit precision, beyond the 8-bit precision widely used in DSP chips for optical communications~\cite{zhang22Optica}. The first full integration of the photonic neural network comprising MRR weight bank and photonic neuron was demonstrated in\cite{tait19PRApplied}. Current silicon photonic foundry chips lack on-chip lasers and optical amplification to support neuron-to-neuron cascadability. The current demonstration uses externally coupled lasers as neuron pumps and passive on-chip front-end impedance connecting the photodetector and MRR modulator to provide neuron-neuron gain and compensate for the circuit losses, at the cost of speed reduction and power consumption increase caused by the on-chip impedance~\cite{huang21NE},\cite{ashtiani22Nature,bandyopadhyay22arXiv}. Other advanced but less mature technologies promise to solve the tradeoffs. As discussed in~\cite{de2019noise}, integrating a transimpedance amplifier with a silicon photonic chip can provide a significant electrical gain to retain cascadability without losing speed. Meanwhile, III/V and heterogeneous III/V-silicon integration platforms promise on-chip lasers and optical amplification~\cite{roelkens2010iii,stabile2021neuromorphic,xiao2021large,nahmias2020laser,peng21arXiv}. The processing latency of this network is determined by the light traveling time in the recurrent loop, which includes the delays at the MRR weight banks, the PD, and the MRR modulator neuron caused by their limited bandwidth and the delay in the waveguide. By using high-speed optoelectronic devices, the processing latency of the photonic RNN is usually smaller than a clock cycle of a DSP chip \cite{nahmias19JSTQE}. In addition, photonic RNNs don't always suffer from the constant data movement, memory fetch and clock distribution operations required in digital RNNs. As a result, the photonic RNN has much lower latency than DSP.

\section{Training Photonic RNN}
\label{sec: Training Photonic RNN}

\subsection{Principle}
The principle of training the photonic RNN is to digitize the continuous photonic RNN model to conventional RNN model, and then the photonic RNN can be trained using the standard deep learning algorithm, as shown in Figure~\ref{fig:framework}. The photonic RNN can be modelled by a set of coupled ordinary differential equations modified from~\cite{tait17SR,peng21arXiv}:

\begin{subequations}\label{eq: Photonic equation}
\begin{center}
    \begin{align}
        \frac{d\bm s(t)}{dt} = \bm{W_{rec}}\bm y(t& - \Delta T)+\bm{w_{in}u}(t) + \bm{b} -\frac{\bm s(t)}{\tau} \\
        \bm y(t) & = \sigma(\bm s(t))
    \end{align}
\end{center}
\label{eq}
\end{subequations}
where $\bm{u}$ and $\bm{y}$ are the input and output WDM signals of the photonic RNN. In our case, both $\bm{u}$ and $\bm{y}$ have $N$ elements, each representing the time series of the distorted and recovered signal, respectively, at a particular wavelength. $N$ is the channel number of the WDM transmission system. $\bm{s}$ and $\bm{b}$ are vectors, representing the neuron state and bias respectively, $\tau$ is the time constant of the photonic neuron. $\Delta T$ is the time delay caused by the feedback optical waveguide. In this paper, we choose $\Delta T$ = $1/B_{data}$, where $B_{data}$ is the signal's baud rate, to 
ensure that the memory of the photonic RNN is approximate to the ratio between dispersion-induced pulse spreading and the original pulse width. $\sigma (\cdot)$ is the nonlinear activation function of the photonic MRR neuron, which is typically a Lorenz-shape function~\cite{tait19PRApplied}. Here we use $\sigma (x)= -0.28/((x-0.5)^2+0.49^2)+1.35$ experimentally characterized from our silicon photonic neuron device~\cite{huang21NE}. Other activation functions can be assembled using coupled-cavity devices~\cite{huang2020chip,jha20OL,campo22JSTQE}. $\bm{W_{rec}}$ is the photonic recurrent weight matrix, and $\bm{w_{in}}$ is the input weight vector. 

The task of training the photonic RNN is to learn the optimized $\bm{W_{rec}}$, $\bm{w_{in}}$ and $\bm{b}$ such that the outputs $\bm{y}$ are the recovered WDM signals. To train the analog photonic RNN, we first discretize Equation~\ref{eq: Photonic equation} using the forward-Euler method, as proposed by~\cite{peng21arXiv}. The discretized photonic RNN model now can be represented by a series of RNN cells similar to software-based RNNs as shown in Figure~\ref{fig:framework}(a), consisting to a series of photonic RNN cells illustrated in Figure~\ref{fig:framework}(b). Figure~\ref{fig:framework}(b) shows a $3\times 3$ recurrent structure as an example. $N\times N$ photonic RNN can be achieved using the same architecture with an $N\times N$ network. In this way, the photonic RNN can be constructed by Pytorch nn.module~\cite{paszke19NIPS} and trained using the back propagation through time (BPTT) algorithm~\cite{werbos90Proc}, similar to software-based RNNs.

\subsection{Training photonic RNN for WDM transmission system}

We use VPI Transmission Maker~\cite{maker6vpi} to simulate the WDM transmission system and generate the dataset. Here we investigate a WDM PAM4 direct-detection system for short-reach transmission. In our simulation, we assume the standard single-mode fiber (SSMF) with a dispersion parameter $D$ = 16 ps/nm/km, a dispersion slope of $D_s =$ 0.08 ps/nm$^2$/km, a nonlinear index $= 2.6 \times$10$^{−20}$ m$^2$/W, an attenuation loss $= 0.2$ dB/km. The fiber loss is compensated by a pre-amplifier with a noise figure of 4 dB. We first look at a 2-channel 56 GBaud transmission system with the signal carrier frequency at 193.1 and 193.2 THz respectively, over a 20 km transmission distance. We generate 102,400 symbols for each channel at the transmitter using pseudorandom bit sequences (PRBSs) with different seeds and obtain 102,400 received symbols from the VPI simulation. 80$\%$ of the symbols are used for training and validation, and 20$\%$ are used as the testing dataset. For $N$-channel WDM system with training sequence length of $M$ per channel, the training objective is to minimize the average mean square error (MSE) loss between the obtained and the expected output of the training input sequences, defined as
$\frac{1}{N}\frac{1}{M}\sum_{i=1}^{N}\sum_{j=1}^{M}[y_{\lambda i}(t_j)-\widehat{y}_{\lambda i}(t_j)]^2$. 
We use Kaiming initialization~\cite{he2015ICCV} to initialize the parameters of the PRNN layer and Xavier initialization~\cite{glorot10JMLR} for output linear regression layer. Adam optimizer is set with an initial learning rate of 0.001, which decays every 1,000 iterations by 0.5~\cite{kingma2014arXiv}. To validate that our network is reliable, we use Monte-Carlo cross-validation with 5 iterations~\cite{xu01CILS}. With this method, we obtain 5 bit error rate values by counting the number of bit errors in total in each iteration. We then obtain the average BER to evaluate the equalization performance. Figure~\ref{fig:framework}(c) shows the mean loss (i.e., MSE) and BER during the training stage. The results are evaluated using the mean BER of the test dataset in different channels. 

\section{Results}

\subsection{Equalization Performance}

\begin{figure}[ht]
\centering
\includegraphics[width=\linewidth]{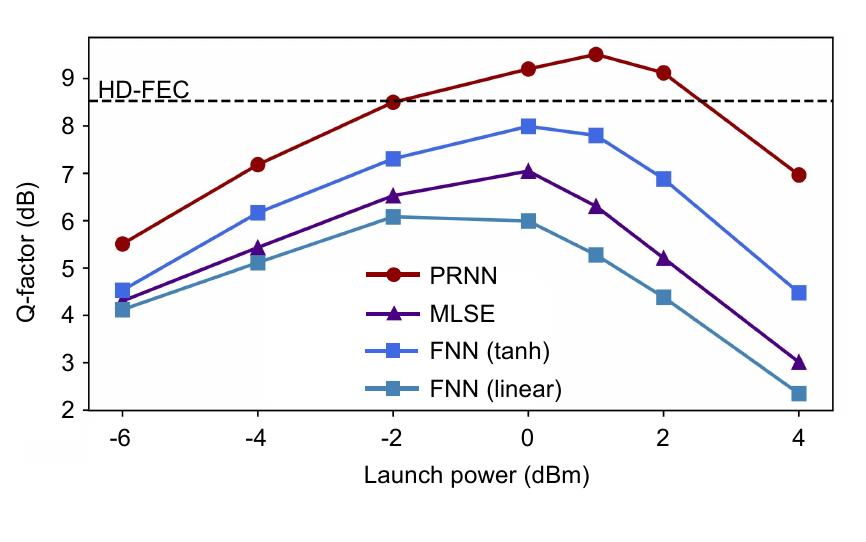}
\caption{Comparison of PRNN and various DSP methods in 2-channel 56 GBaud/$\lambda$ WDM PAM4 signals over 20 km transmission link. PRNN: photonic recurrent neural network; FNN: feed-forward neural network; MLSE: maximum likelihood sequence estimation. tanh: hyperbolic tangent activation function.}
\label{fig:DSP Comparison}
\end{figure}

After training, we test the photonic RNN's ability to resolve fiber-induced distortions, particularly fiber nonlinearity, in a 2-channel 56 GBaud/channel transmission link over 20 km distance. We compare the photonic RNN with different DSP methods including, maximum likelihood sequence estimation (MLSE) and feed-forward neural network equalizer (FNN), as shown in Figure~\ref{fig:DSP Comparison}. MLSE is a linear equalizer commonly used for dispersion compensation~\cite{agazzi05JLT}. The MLSE used in the comparison is operated with two samples per symbol and a memory length of 8 symbols according to the ratio between dispersion-induced pulse spreading and the original pulse width. The coefficients of MLSE are updated to minimize the MSE. The FNN design is similar to that published in~\cite{ranzini21JLT}, having an input layer accepting 8 time-delayed symbols, one hidden layer with 16 neurons, and an output layer with 2 neurons outputting the recovered signals. The input size is approximate to the dispersion-induced pulse spreading divided by the original pulse width. To study the performance of linear and nonlinear equalizers, we include two FNNs in our comparison, one with hyperbolic tangent ($tanh$) activation function and the other without activation functions (i.e., a linear network). It is worth noting that the DSP algorithms can only be implemented on a channel-by-channel basis, while the photonic RNN is able to process all WDM channels simultaneously.

Figure~\ref{fig:DSP Comparison} shows the results of system Q-factor performance as a function of optical powers launched into the fiber input under different compensation schemes. The Q-factor value is calculated by~\cite{freude12ICTON}:
\begin{equation}
  Q = 20\cdot\log_{10}(\sqrt{2}\text{erfc}^{-1}(2\text{BER}))
\end{equation}
where BER is the average BER over all WDM channels.  As shown in Figure~\ref{fig:DSP Comparison}, 
we see that the photonic RNN significantly outperforms all other DSP-based compensation algorithms due to its capability of processing both intra- and inter-channel distortions. The optimum launched power is 1 dBm with a BER of 1.4$\times$10$^{-3}$ (corresponding to 9.51 dB Q-factor). FNN with $tanh$ activation functions is only capable of mitigating intra-channel distortions including dispersion and self-phase modulation, thus the optimum BER after compensation is 6.2$\times$10$^{-3}$ (7.96 dB Q-factor) at a launch power of 0 dBm, still worse than the hard-decision forward error correction (HD-FEC) threshold (3.8$\times$10$^{-3}$). The FNN without nonlinear activation function and MLSE are linear equalizers, so they can only compensate for fiber dispersions and have the worst BER performances.

\begin{figure}[ht]
\centering
\includegraphics[width=\linewidth]{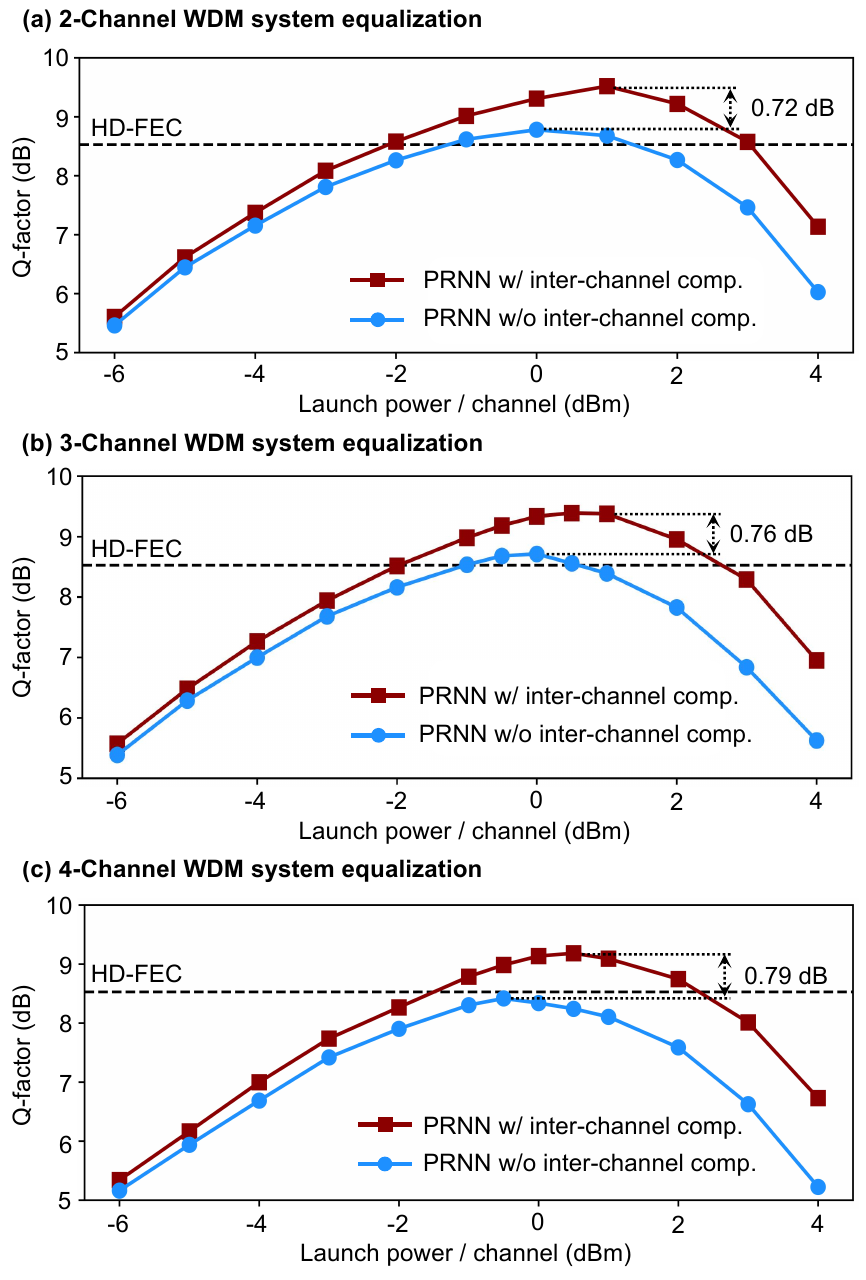}
\caption{Equalization performance for (a) 2, (b) 3, and (c) 4 channels WDM systems respectively. The WDM systems have 56 GBaud/$\lambda$ PAM4 signals up to 20 km transmission with different launch power per channel. The red and blue curves represent the average equalization performance of all channels using PRNN with and without inter-channel nonlinearity compensation, respectively. PRNN: photonic recurrent neural network.}
\label{fig:interchannel}
\end{figure}

To further confirm that the photonic RNN can compensate for inter-channel nonlinearity, we conduct the following simulations based on the same photonic RNN. In two cases, the transmission link contains 2 $\times$ 56 GBaud PAM4 signals. In one case shown in the blue curve in Figure~\ref{fig:interchannel}(a), we train the neural network using the training dataset of channel 1 and evaluate the BER performance of channel 1. Due to lacking the data from the second channel, this case can only address intra-channel distortions similar to the FNN. In the other case, the photonic RNN is trained using the training datasets from both channels, as shown by the red curve in Figure~\ref{fig:interchannel}(a). In this case, since the neural network can learn the response of all WDM channels rather than only a single channel, the inter-channel crosstalk can be compensated. As a result, we observe a 0.72 dB improvement in the Q-factor and a 1 dB improvement in optimum launch power than in the first case.

\subsection{Channel scalability}  \label{subsec: scalability}





We investigate the performance of the proposed photonic RNN with more WDM channels. The signal speed is 56 GBaud/s/channel and the transmission distance is 20 km. The channel space is 100 GHz. The results of 3 and 4 channels is shown in Figure~\ref{fig:interchannel}(b)(c). Similar to Figure~\ref{fig:interchannel}(b)(c), the red curve shows the results of training all channels simultaneously to compensate for inter-channel crosstalk, whereas the blue curve shows the results of training only one channel. In the 3-channel system, the optimal launch power increases from 0 dBm to 1 dBm, and the optimal Q factor increase from 8.63 dB to 9.39 dB. In the 4-channel system, the optimal launch power increases from -0.5 dBm to 0.5 dBm, and the optimal Q factor increase from 8.38 dB to 9.17 dB.




We find it becomes more challenging to train more channels because the current objective function is difficult to render optimized BER performances for every channel simultaneously. The WDM signal in one channel is more likely to be influenced by the signals from its adjacent channels than by the channels farther away. Thus in principle, the weights between neighboring channels should be more significant than distant channels. However, the current training objective is the MSE averaged across all channels. The consequence of this non-optimized objective function is the trained weights between distant channels are larger than they should be. A potential solution to this problem is to assign a dedicated objective function to every individual channel~\cite{chen22arXiv}.

\subsection{Data rate-distance product comparison with other published work} 

\begin{figure}[ht]
\centering
\includegraphics[width=\linewidth]{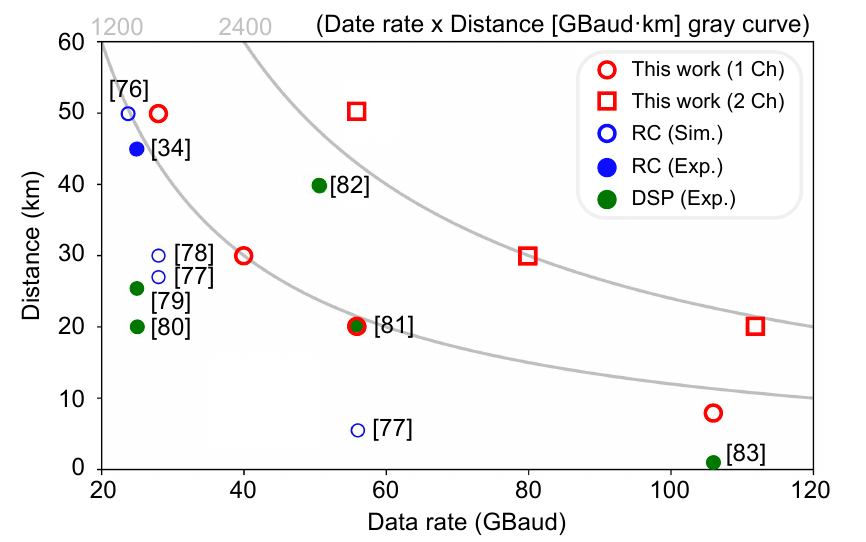}
\caption{Comparison of different PAM4 equalization systems with reservoir computing (RC)~\cite{bogris20JSTQE, argyris18SR, argyris2019pam,yu21OECC} and DSP~\cite{xu21PTL,xu20JLT, tang20JLT,yu20OE,sang21ECOC} in terms or data rate and distance. Ch: channel; Sim: simulation; Exp: experiment.
}
\label{fig: RC DSP comparison}
\end{figure}


To further evaluate the capability of our proposed photonic RNN, we simulate the transmission performance under different single-lane data rates to find out the maximum transmission distance at BER = 3.8$\times 10^{-3}$. Figure~\ref{fig: RC DSP comparison} shows the data rate-distance product and the comparison between our work (red dots) and other published transmission system demonstrations with PAM4 signals. We have included both DSP (green dots) and photonic reservoir computing (blue dots) in the comparison. The gray curves are the contour lines given by the product of the data rate and the distance. When only considering a single channel, our work can achieve a data rate-distance product of 1200 GBaud$\cdot$km, outperforming most of the other approaches except~\cite{yu20OE}. The algorithm used in~\cite{yu20OE} requires a feedforward equalizer and MLSE. Its problem is that in the system with more distinct dispersion-induced pulse broadening, the computational complexity increases enormously. In contrast, our approach only requires a very simple photonic RNN with one photonic neuron in the single-channel transmission system. When there are two channels in the transmission system, we see that our works' data rate-distance products (red squares) are approximately doubled to 2400 GBaud$\cdot$km, implying that our approach can extend the transmission performance by compensating for the inter-channel crosstalk in multi-channel WDM system. Other works under comparison only consider single transmission channel.

\section{Power Consumption and Latency Analysis}
\label{subsec: power consumption}

This section will show that photonic RNNs not only exceed DSP in channel equalization performance but they also have the potential to drastically reduce power consumption and latency in optical communication systems. When estimating the power consumption and latency of PRNN, we assume that the system uses devices accessible in commercial silicon photonics foundries. For the DSP, we assume that it uses 5 nm CMOS technology. We realize the challenges of quantifying the power consumption and latency of DSP at the chip level. Thus, we only focus on the basic multiplication and addition operations when analyzing the DSP. When analyzing the photonic RNN, we have included fundamental sources of on-chip energy dissipation that are needed for a photonic neural network to operate (such as weight banks, photonic neurons, pump lasers, as well as most of the peripheral circuits (DACs)).

\subsection{Power Consumption}

\subsubsection{Photonic RNN with Silicon Photonic Platform}

We follow the outline of power consumption of photonic neural network derived in previous publications~\cite{tait19PRApplied,huang21NE, huang22JLT, nahmias19JSTQE}. The power consumption of a $N\times N$ photonic RNN is given as:
\begin{equation}\label{eq: Photonics power}
\begin{aligned}
    P_{Photonics}&=\underbrace{[(E_{dac}+E_{mrr})\cdot B_{data} +P_{driver} + P_{heater}]\cdot N^2}_{\rm Weighting~power} \\
    &+ \underbrace{[(E_{rec} +E_{mod})\cdot B_{data} +P_{pump}+ P_{heater}]\cdot N}_{\rm Neurons~power} 
\end{aligned}
\end{equation}
The first term in Equation~\ref{eq: Photonics power} is the power consumption cost by photonic weights, where $B_{data}$ is the baud rate, $E_{dac}$ and $E_{mrr}$ are the energy consumption spent on digital-to-analog converters and optical modulators respectively, to actuate weights. High-speed DACs based on 7-nm FinFET~\cite{groen20JSSC} are used to convert the digital weights to analog voltages. $P_{driver}$ is the power consumption of the MRR driver. Microheaters are used to tune the MRR resonance to match the pump wavelength. Thus, $P_{heater}$ is the power consumption of the microheater and its control circuit for resonance alignment. The second term is the power consumption of photonic neurons. In the physical RNN, the signal power ratio between the input and output in the recurrent loop must be greater than unity~\cite{tait19PRApplied}. For photonic neurons, physical cascadability is expressed as the input and output being both optical, and their modulation depth and optical power remain unity. Many factors account for losses inside the closed loop, including (1) optical losses of optical devices and waveguides, which can be compensated by an external and powerful enough optical pump, and (2) electrical losses due to inefficient modulation depth. In our analysis, we assume a receiver with photodetector following transimpedance amplifier (TIA)~\cite{li20OE}. The electrical loss is assumed to be compensated by a TIA between the O/E and E/O stages of photonic neurons. The TIA amplifies the photocurrent to drive the following modulator in full swing while sidestepping the gain-bandwidth trade-off in the neuron unit. $E_{rec}$ and $E_{mod}$ represent the energy consumption of electrical receiver (including the photodetector and TIA) and the MRR modulator neuron, respectively. The optical pump not only serves as an optical carrier of neuron output, but it also injects extra optical energy into the neuron to compensate for optical losses due to insertion losses and limited extinction ratio.The network needs to ensure greater-than-unity O/E/O efficiency. Thus, the neuron pump must provide sufficient photocurrent (after detection) to drive the neuron modulator in full swing. It also needs to compensate for various losses in the chip, such as coupling loss and device insertion losses. The $P_{pump}$ used in our calculation is modified from~\cite{tait19PRApplied} by further considering the losses in the peripheral circuits. The value of power consumption would depend on different technology developments~\cite{huang21NE}. Similar to~\cite{huang21NE,huang22JLT}, in this paper, we assume that the system uses the silicon PN modulators and Ge-on-Si photodetectors available in commercial silicon photonics foundries. Table~\ref{table:Modeling parameters} in Appendix~\ref{sec: parameters} summarizes the parameter values used to calculate the power consumption of the photonic neural network.

\subsubsection{DSP}
We follow the paper by Pillai $\textit{et al}.$~\cite{pillai14JLT} to calculate the power consumption of DSP. Generally, the power consumption in DSP is the number of channel $N$ times the sum of the energy of the required operations $E_{op}$ multiplied by how many operation are required in a second $N_{op}$, as shown in is Equation~\ref{eq: DSP power}. 

\begin{equation}\label{eq: DSP power}
\begin{aligned}
    P_{DSP}=N \cdot \sum E_{op}N_{op}
\end{aligned}
\end{equation}
For simplicity, we exclude the power consumption of serializer/deserializer, which although is a highly challenging and power consuming module in DSP, particularly for high speed signals. We look at two algorithms, FNN and MLSE, and focus on their number of operations $N_{op}$, especially multiplication $N_M$ and addition $N_A$ which account for most power consumption. 

We first consider the $I\times H\times O$ 3-layer delayed FNN system for single channel compensation, where $I$, $H$, $O$ represent the number of neurons in each layer. Assuming that the FNN processes data sequentially, the number of operations completely depends on the network size and channel number. The number of multiplication $N_{M,fnn}$ and addition $N_{A,fnn}$ operations is as follows:
\begin{subequations}\label{eq: FNN}
\begin{center}
    \begin{align}
       N_{M,fnn}&=(I\cdot H+H\cdot O)\cdot B_{data} \\
       N_{A,fnn}&=((I-1)\cdot H+(H-1)\cdot O)\cdot B_{data}
    \end{align}
\end{center}
\label{eq}
\end{subequations}


We then adapt MLSE structure from~\cite{yu20OE} to derive the operations number of multiplication $N_{M,mlse}$ and addition $N_{A,mlse}$. In order to search for the most probable signal sequence using MLSE, the memory length must scale with the dispersion-induced pulse spreading length which leads to the exponentially increased operation number~\cite{yu20OE}, as shown in follows: 

\begin{subequations}\label{eq: MLSE}
\begin{center}
    \begin{align}
       N_{M,mlse} &=L\cdot P^{L+1}\cdot B_{data} \\
       N_{A,mlse} &=(L+2)\cdot P^{L+1}\cdot B_{data}
    \end{align}
\end{center}
\label{eq}
\end{subequations}
where $L=4$ is the truncated channel length and $P=2$ is the number of symbol to construct a reduced-state trellis~\cite{yu20OE}. 



\begin{figure}[ht]
\centering
\includegraphics[width=\linewidth]{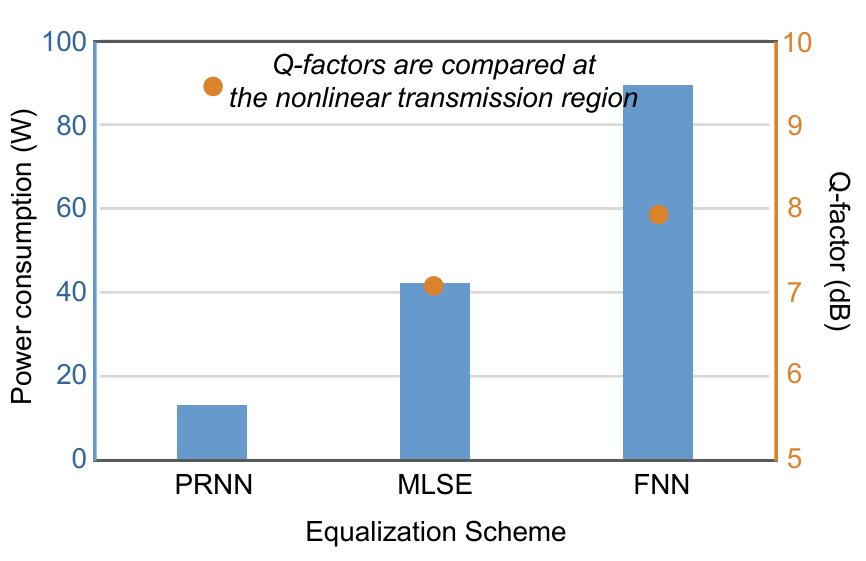}
\caption{Power consumption of 1.6 T transceiver module and Q-factor of 2-channel 56 GBaud/$\lambda$ WDM PAM4 signals over a 20 km transmission link with different equalization schemes. PRNN: photonic recurrent neural network; MLSE: maximum likelihood sequence estimation; FNN: feed-forward neural network. In calculating the power consumption of PRNN, we
assume that the system uses the silicon PN modulators and Ge-
on-Si photodetectors available in commercial silicon photonics
foundries. MLSE and FNN are DSP approach using 5 nm CMOS node.
}
\label{fig:Power Comparison}
\end{figure}

Figure~\ref{fig:Power Comparison} plots the predicted power consumption of the next generation 1.6 T transceiver for data center communications with 8 WDM channels $\times$ 200 Gbit/s/lane. $N = 8$ in Equation~\ref{eq: DSP power}. We assume 5 nm CMOS node is used for DSP chips. The operation energy consumption of multiplication $E_{op,M}$ and addition $E_{op,A}$ per information bit are 347.5 fJ/b and 43.4 fJ/b, respectively, according to~\cite{pillai14JLT}. We can see from Figure~\ref{fig:Power Comparison} that the PRNN power consumption is predicted to be 12.7 W which is lower than other DSP algorithms while achieving the best signal quality. In comparison, the estimated power consumption of MLSE is 42.3 W, and the FNN-based equalizer has a power consumption of over 90.0 W, both are beyond the thermal management capacity of today’s semiconductor packaging technology~\cite{QSFPDDWhitepaper}. The parameters used to generate Figure~\ref{fig:Power Comparison} are summarized in Table~\ref{table:Modeling parameters} of Appendix~\ref{sec: parameters}.

\subsection{Latency}

The processing latency of this network is determined by
the light traveling time in the recurrent loop, which includes
the delays at the MRR weight bank, the PDs, and the MRR modulator neurons caused by their limited bandwidth and the delay in the waveguide. The MRRs in the weight bank have a low Q-factor of only a few thousand, and thus the weights are computed in a single time step within a few picoseconds. The delay in the photodetector and MRR modulator neuron is approximately equal to the inverse of their total bandwidth (i.e., $1/B_{mod}+1/B_{pd}$). The delay in the recurrent waveguide equals one symbol duration (i.e., $1/B_{data}$). To compensate for the dispersion, the signals need to recirculate in the loop for at least $M$ times. $M$ is approximately the ratio between dispersion-induced pulse spreading and the original pulse width. Therefore, the total latency of PRNN is $M(1/B_{mod}+1/B_{pd})+(M-1)(1/B_{data})$~\cite{nahmias19JSTQE}.  $M=8$ in the 1.6 T transmission system with 8 PAM4 WDM channels and 100 Gbaud/lane. Therefore, the total latency of a photonic RNN is approximately 470~ps.

A significant difference between DSP and photonic systems is that the latency of DSP systems is fundamentally bounded by the system clock $f_{clock}$. The DSP clock $f_{clock}$ is usually set at a few hundreds of MHz in most DSP systems for the best energy efficiency\cite{pillai14JLT}. It is challenging to estimate the latency in DSP because the latency is not only determined by the required operations but also the exact circuit layout. Thus for simplicity, we consider the best scenario for DSP where DSP hardware can always operate at line rate speed via massive parallelism, so only the delay in serial to parallel (s/p) conversion is included. The latency of s/p conversion is $2B_{data}/f_{clock}^2$~\cite{huang22JLT}. With the parameters values in Table~\ref{table:Modeling parameters}, we can obtain the minimum latency of DSP processor's latency of 1.6 $\mu$s, which is more than 3,400 times longer than 470 ps in photonic RNN.

\section{Conclusion}
\label{sec:conclusion}


In this paper, we present a photonic recurrent neural network (RNN) compatible with the silicon photonic platform for intra and inter-channel impairments compensation in WDM systems. We demonstrate in simulation that our photonic RNN equalizer can outperform digital MLSE and FNN equalizer in a 56 Gbaud WDM PAM4 transmission system in terms of optimum Q-factor and launch power. A higher baud rate can be achieved by replacing the optical modulators and photodetectors in the photonic RNN with those with matched speeds. Due to inter-channel nonlinearity compensation, our photonic RNN improves the signals' Q-factor by 0.72, 0.76, and 0.79 dB in 2, 3, and 4 channels WDM systems, respectively, compared to other schemes without inter-channel compensation capability. Furthermore, we show that our photonic RNN  has a better data rate-distance product than most other channel equalization using photonic reservoir computing and DSP, except for some computational expensive DSP algorithms. In addition to the advantages in BER performance, our photonic RNN has a unique capability of processing WDM communication signals of multiple channels directly in the optical domain without ADCs. This capability ensures the photonic RNN has much greater energy efficiency compared to other photonic neural network and DSP approaches. Our proposed photonic RNN has shown significantly reduced power consumption and processing latency compared to the conventional DSP hardware when using the mature silicon photonic foundry technologies. Prior research has suggested that the power consumption can be further reduced by two orders of magnitude with advanced nanophotonics~\cite{nahmias19JSTQE}.

\appendices

\section{Calculation Parameters Values}
Related values of involved parameters for calculation are listed in Table~\ref{table:Modeling parameters}.
\label{sec: parameters}

\begin{table}[h!]
\renewcommand{\arraystretch}{1.4}
\centering
\caption{Modeling parameters}
\label{table:Modeling parameters}
\begin{tabular}{lp{3cm}p{3cm}p{4cm}}
\hline
 \textbf{Symbol} & \textbf{Quantity} & \textbf{Value} \\ [0.5ex] 
 \hline
 $N$   & Signal channel number       & 8  \\ 
 $B_{data}$ & Signal baud rate   & 100 GBaud/s  \\
 $E_{dac}$ &Energy consumption of 7-nm FinFET DAC & 1.05 pJ/b~\cite{groen20JSSC} \\
 $E_{mrr},E_{mod}$ & Energy consumed to change the voltage on the modulator  &$C_{mod}V_r^2/4$ \\ 
 
 $C_{mod}$ & Modulator capacitance    &30 fF \\
 $V_r$ &Modulator voltage swing &4.8 V \\
 $P_{driver}$  &Power consumed to drive modulator   & 61 mW~\cite{ramon19JLT} \\
 $P_{heater}$ &The power consumption of heater and control circuits for each MRR to resonance alignment & 4.6 mW~\cite{jayatilleka19Optica}\\
 $E_{rec}$ & Optical receiver with TIA & 550 fJ/b~\cite{li20OE} \\
 $P_{pump}$ & Pump laser power & $2V_r/(\pi \eta_{wall-plug} \alpha \beta R_{pd} R_{load})$\cite{tait19PRApplied} + $P_{tec}$ \\
 $\eta_{wall-plug}$ & Wall-plug efficiency of pump lasers & 0.3 \\
  $\alpha$ &  Coupling efficiency between laser and photonic chip  &  0.8~\cite{siew21JLT}   \\
  $\beta$ &  WDM multiplexer efficiency  &  0.83~\cite{tan11OE}  \\
 $R_{pd}$ & Photodiode responsivity  & 1 A/W  \\
 $R_{load}$ & Resistance of the receiver junction & 3 k$\Omega$ \\
 $P_{tec}$ & TEC power for each laser & 1.3 mW~\cite{gao17OQE}\\
 $E_{op,A}$ & Average energy per $n_b$-bit adder op.  & $2.57n_bp_tV^2$ fJ/b  \\
 $E_{op,M}$ & Average energy per $n_b$-bit multiplier op.  & $2.57n_b^2p_tV^2$ fJ/b  \\
 $n_b$  & Average bit resolution of the DSP module &8\\
 $p_t$  & CMOS process technology feature size     &5 nm\\
 $V$    & CMOS supply voltage                      &0.65 V\\
 
 $I,H,O $ & FNN size of different layers & 16, 16, 2\\
 $N_{fnn,M}$ & The number of multiplication in FNN  & 288$\cdot B_{data}$ \\
 $N_{fnn,A}$ & The number of addition in FNN  & 270$\cdot B_{data}$ \\
 $N_{mlse,M}$ & The number of multiplication in MLSE  & $LP^{L+1}B_{data}$ \\
 $N_{mlse,A}$ & The number of addition in MLSE  & $(L+2)P^{L+1}B_{data}$ \\
 $L$ &The truncated channel of length &4~\cite{yu20OE} \\
 $P$ &The number of symbol to construct a reduced-state trellis & 2~\cite{yu20OE} \\
 $B_{mod}$ & Modulator bandwidth    & 40 GHz \\
 $B_{pd}$ & Photodiode bandwidth    & 40 GHz \\
 $f_{clock}$ & Clock frequency of DSP    & 500 MHz \\


\hline
\end{tabular}
\end{table}


\section*{Acknowledgment}

This work was supported by CUHK Direct Grant 170257018, RNE-p4-22 of the Shun Hing Institute of Advanced Engineering, CUHK, and CUHK startup fund. The devices were fabricated at the IME ASTAR foundry in Singapore. Fabrication support was provided via the Natural Sciences and Engineering Research Council of Canada (NSERC), Silicon Electronic-Photonic Integrated Circuits (SiEPIC) Program and the Canadian Microelectronics Corporation (CMC). 

\ifCLASSOPTIONcaptionsoff
  \newpage
\fi

\input{main.bbl}

\bibliographystyle{IEEEtran}

\listoffixmes
\begin{IEEEbiographynophoto}{Benshan Wang}
received the B.Eng. degree in electrical information engineering from Wuhan University, Wuhan, China, in 2020. He is currently
working toward the Ph.D. degree at the Chinese University of Hong Kong, Hong Kong. His research interests include neuromorphic photonics, photonic integrated circuits, optical signal processing, and nonlinear dynamics of photonic systems for neuromorphic applications.
\end{IEEEbiographynophoto}

\begin{IEEEbiography}[{\includegraphics[width=1in,height=1.25in,clip,keepaspectratio]{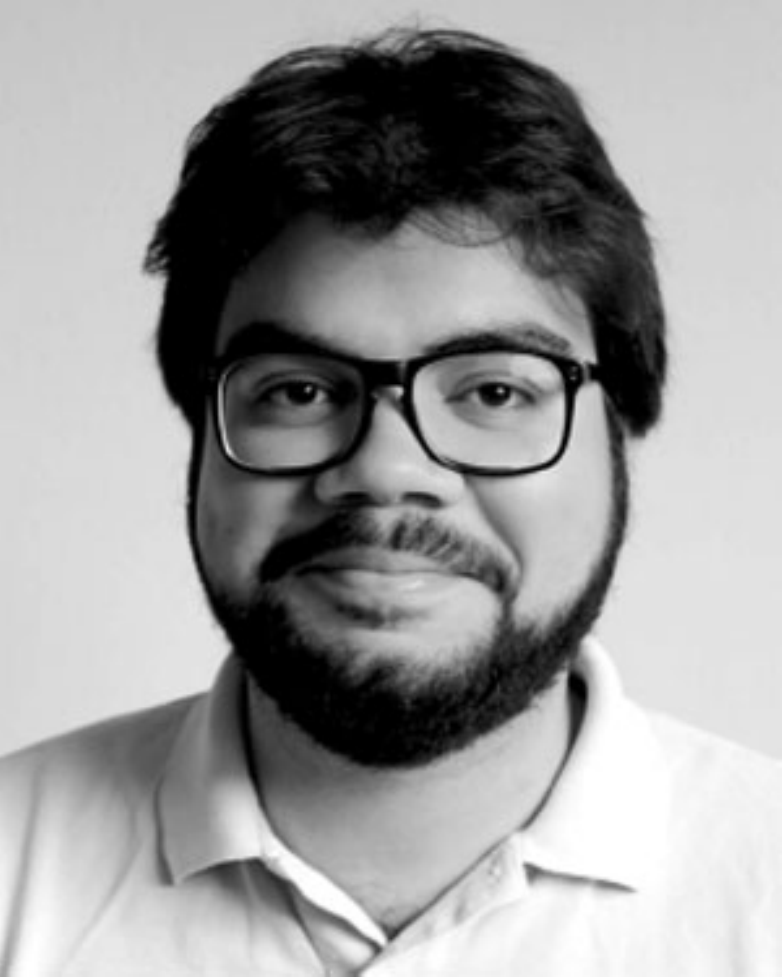}}]{Thomas Ferreira de Lima}
received the B.Sc. and the Ingénieur Polytechnicien master’s degrees from Ecole Polytechnique, Palaiseau, France, in 2016 with a focus on physics for optics and nanosciences. He received the Ph.D. degree in electrical engineering in 2022 with the Lightwave Communications Research Laboratory, Department of Electrical Engineering, Princeton University, Princeton, NJ, USA. He is currently a Researcher with the NEC Laboratories America, Inc., Princeton, NJ, USA. His research interests include: real-time integrated photonic systems, nonlinear signal processing with photonic devices, spike-timing based processing, ultrafast cognitive computing, and dynamical light-matter neuro-inspired learning and computing.
\end{IEEEbiography}

\begin{IEEEbiographynophoto}{Bhavin Shastri}
(Senior Member, IEEE) received the Honors B.Eng. (with distinction), M.Eng., and Ph.D. degrees in electrical engineering (photonics) from McGill University, Montreal, QC, Canada, in 2005, 2007, and 2012, respectively. He is currently an Assistant Professor of engineering physics with Queen’s University, Kingston, ON, Canada, and a Faculty Affiliate with the Vector Institute for Artificial Intelligence, Canada. He was an NSERC and Banting Postdoctoral Fellow (2012–2016) and an Associate Research Scholar (2016–2018) with Princeton University. He has authored or coauthored more than 70 journal articles and 100 conference proceedings, seven book chapters, and given more than 60 invited talks and lectures including five keynotes and three tutorials. His research interests include silicon photonics, photonic integrated circuits, neuromorphic computing, and machine learning. He is a co-author of the book, \textit{Neuromorphic Photonics} (Taylor Francis, CRC Press, 2017), a term he helped to coin. Dr. Shastri was the recipient of the 2022 SPIE Early Career Achievement Award and the 2020 IUPAP Young Scientist Prize in Optics for his pioneering contributions to neuromorphic photonics from the ICO. He is a Senior Member of Optica (formerly OSA), 2014 Banting Postdoctoral Fellowship from the Government of Canada, the 2012 D. W. Ambridge Prize for the top graduating Ph.D. student at McGill, an IEEE Photonics Society 2011 Graduate Student Fellowship, a 2011 NSERC Postdoctoral Fellowship, a 2011 SPIE Scholarship in Optics and Photonics, a 2008 NSERC Alexander Graham Bell Canada Graduate Scholarship, including the Best Student Paper Awards at the 2014 IEEE Photonics Conference, 2010 IEEE Midwest Symposium on Circuits and Systems, the 2004 IEEE Computer Society Lance Stafford Larson Outstanding Student Award, and the 2003 IEEE Canada Life Member Award.
\end{IEEEbiographynophoto}

\begin{IEEEbiographynophoto}{Paul R. Prucnal}
(Life Fellow, IEEE) is a Professor of Electrical Engineering at Princeton University. He is best known for his seminal work in Neuromorphic Photonics, optical code division multiple access (OCDMA) and the invention of the terahertz optical asymmetric demultiplexer (TOAD). Prucnal received his A.B. In mathematics and physics from Bowdoin College, graduating \textit{summa cum laud}. He then earned M.S., M.Phil. and Ph. D. degrees in electrical engineering from Columbia University. After his doctorate, Prucnal joined the faculty at Columbia University in 1979. As a member of the Columbia Radiation Laboratory, he performed groundbreaking work in Optical CDMA, which initiated a new research field in which more than 1000 papers have since been published, exploring applications ranging from information security to communication speed and bandwidth. In 1988, he joined the faculty at Princeton University. Prucnal is author of the book, \textit{Neuromorphic Photonics}, and editor of the book, \textit{Optical Code Division Multiple Access: Fundamentals and Applications}. He has authored or co-authored more than 300 journal articles, 390 conference papers and 34 book chapters, and holds 28 U.S. patents. He is a fellow of the Institute of Electrical and Electronics Engineers (IEEE), the Optical Society of America (OSA) and the National Academy of Inventors (NAI), and a member of Phi Beta Kappa and Sigma Xi. He was the recipient of the Gold Medal from the Faculty of Mathematics, Physics, and Informatics at Comenius University, for leadership in the field of optics and numerous teaching awards at Princeton, including the President’s Award for Distinguished Teaching, the Distinguished Teacher Award, School of Engineering and Applied Science, the Engineering Council Lifetime Achievement Award for Excellence in Teaching, the Graduate Mentoring Award, the Walter Curtis Johnson Prize for Teaching Excellence.
\end{IEEEbiographynophoto}

\begin{IEEEbiography}[{\includegraphics[width=1in,height=1.35in,clip,keepaspectratio]{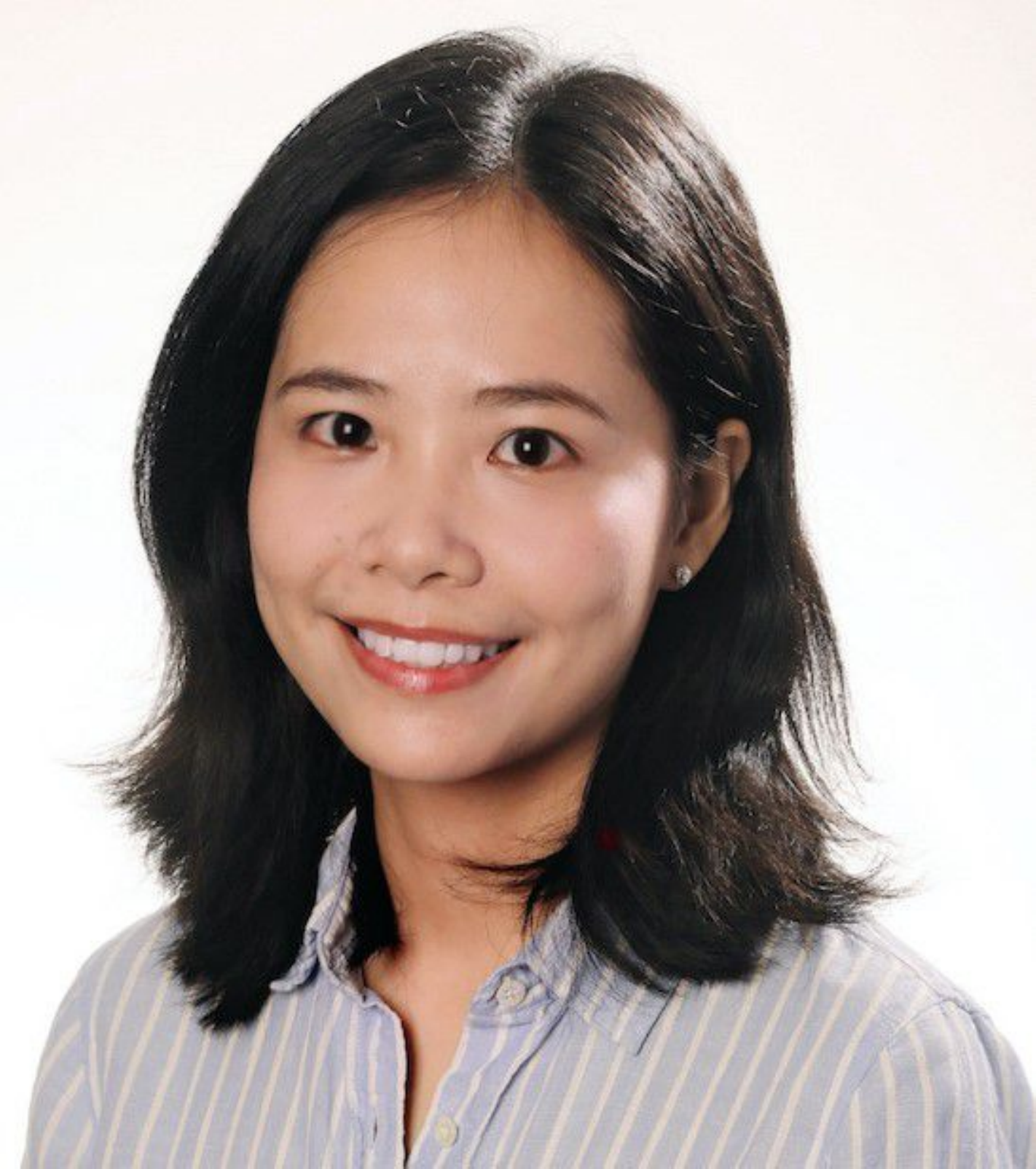}}]{Chaoran Huang}
received the B.Eng. degree from the Huazhong University of Science and Technology, Wuhan, China, in 2012, and the Ph.D. degree from the Chinese University of Hong Kong, Hong Kong, in 2016. She is currently an assistant professor at the Chinese University of Hong Kong (CUHK). Before joining CUHK, she was a Postdoctoral Research Fellow with Princeton University, Princeton, NJ, USA, from 2017 to 2021. She has authored more than 40 peer-reviewed research papers, three book chapters, and one US patent, and has given more than 15 invited talks at different international conferences. Her research interests include photonic neuromorphic computing, optical computing, silicon photonics, photonic integrated circuits, and optical communications. Her current research interest is to develop neuromorphic photonic platforms, including novel devices, photonic integrated circuits, and complementary algorithms for high-performance AI computing and information processing. She has been the TPC Member of several international conferences and is a frequent reviewer for different journals in IEEE, OSA, and the Nature Publishing Group. She was the recipient of 2019 Rising Stars Women in Engineering Asia, and was nominated by Princeton University to compete for Blavatnik Regional Awards for Young Scientists.
\end{IEEEbiography}



\end{document}